\documentstyle[prl,aps,twocolumn,epsfig]{revtex}

\begin{document}
\preprint{to be submitted to Phys. Rev. B}

\twocolumn[\hsize\textwidth\columnwidth\hsize\csname@twocolumnfalse\endcsname
\draft

\title{
Determination of the complex microwave photoconductance\\
of a single quantum dot
}

\draft
\author{
H.~Qin$^{1}$, F.~Simmel$^{1,\dag}$,
R.~H.~Blick$^1$, J.~P.~Kotthaus$^1$,\\ W.~Wegscheider$^{2,\ddag}$, and
M.~Bichler$^2$
}

\address{
$^1$Center for NanoScience and Sektion Physik,
Ludwig-Maximilians-Universit\"at,\\ Geschwister-Scholl-Platz 1,
80539 M\"unchen, Germany\\
$^2$Walter-Schottky-Institut der Technischen Universit\"at M\"unchen,\\
Am Coulombwall, 85748 M\"unchen, Germany
}

\date{\today}
\maketitle

\begin{abstract}
A small quantum dot containing approximately 20 electrons is
realized in a two-dimensional electron system of an AlGaAs/GaAs
heterostructure.
Conventional transport and microwave spectroscopy reveal 
the dot's electronic structure. 
By applying a coherently coupled two-source technique, we are able to 
determine the complex microwave induced tunnel current. 
The amplitude of this photoconductance resolves photon-assisted tunneling~(PAT) 
in the non-linear regime through the ground state and an excited state as well. 
The out-of-phase component~(susceptance) allows to 
study charge relaxation within the quantum dot 
on a time scale comparable to the microwave beat period. \\

\end{abstract}
]

Spectroscopy on quantum dots is commonly performed either by
non-linear transport~\cite{weis93,kouwenhoven97} or by microwave
measurements~\cite{kouwenhoven94,blick95,oosterkamp97,oosterkamp98}.
Ordinary linear transport, i.e.~under a small forward bias
$V_{ds}$ between source and drain contacts and without microwave
irradiation, only involves quantum dot ground states. In the
non-linear case, by applying a finite bias across the 
`artificial atom'~\cite{ashoori96}, also excited quantum dot states can
participate in transport. Alternatively, in the presence of a
microwave field electrons can absorb or emit photons and thus
reach excited quantum dot states otherwise not available in linear
transport -- a phenomenon known as photon-assisted tunneling~(PAT)~\cite{tien63}. 
In a combination of the two methods described, we use two coherently 
coupled microwave sources with a slight frequency offset and detect 
the complex photoconductance signal~(microwave induced tunneling current) at 
the difference frequency. In this way, we are not limited by the 
broadening of the conductance resonances due to the finite bias and 
thus are able to resolve PAT in the non-linear regime as well.
Furthermore, the detected photoconductance contains the 
in-phase part~(conductance) and out-of-phase part~(susceptance). 
The variation of these two different responses indicates 
the different dynamics of the involved transport processes 
through the artificial atom.

For the observation of PAT through excited states the size of the 
quantum dot system is crucial: First, the dot has to be small
enough to have a mean energy level spacing $\bar{\Delta}$ large
compared to the intrinsic or thermal broadening of the conductance
resonances, i.e.~$\Gamma, k_BT < \bar{\Delta}=2\hbar^2/m^* r^2$,
where $\Gamma$ denotes the intrinsic level broadening, $T$ the
temperature, $r$ the radius of the dot and $m^* \approx 0.067m_{e}$ the effective electron mass.
Second, the excited state must be attainable via absorption of one or a few photons,
i.e.~$hf \approx \bar{\Delta}$, where $f$ is the microwave
frequency. In order to form such a small laterally confined
quantum dot, patterned split gates are adopted to selectively
deplete the two-dimensional electron system (2DES) of an
AlGaAs/GaAs heterostructure. The split gates are fabricated on the
surface of the heterostructure using electron beam lithography. A
schematic drawing of the structure is shown in the inset of
Fig.~1. The gate structure separates a small electronic island 
(with a lithographic radius of about 100~nm) from the 2DES 
via tunneling barriers. The 2DES itself is located
50~nm below the surface of the heterostructure and has a carrier
density of $n_s \approx 2 \times 10^{11}$~cm$^{-2}$ and a low temperature
mobility of $\mu \approx 8 \times 10^5$~cm$^2$/Vs.

In order to characterize the electronic structure of the
artificial atom, at first standard direct current (dc) transport
measurements without high frequency irradiation are performed. The
measurements are conducted in a dilution refrigerator at $140$~mK
bath temperature which is higher than its possible minimum value
of $20~$mK. This is due to heat leakage through the coaxial lines
used to couple the high frequency radiation to our sample. In the
tunneling regime at $V_{ds}=0$ the conductance of the quantum dot
is normally zero due to Coulomb blockade
(CB)~\cite{kouwenhoven97}. By tuning one of the gate voltages,
however, the potential of the dot can be varied to align a
discrete quantum dot state with the chemical potentials of the
leads~\cite{sloppy} which results in a conductance resonance. The
gate voltage range over which the CB is lifted can be increased by
applying a finite bias across the quantum dot. Changing gate and
bias voltage simultaneously therefore leads to a diamond-shaped
conductance pattern in the $V_{ds}-V_g$-plane. The result is
displayed in Fig.~1, where the differential conductance
$dI_{ds}/dV_{ds}$ in the vicinity of a resonance is shown as a function
of forward bias $V_{ds}$ and gate voltage $V_g$. For convenience,
the gate voltage is rescaled to $\Delta E = -e \alpha \Delta V_g$, 
which is the energetic distance from the ground state resonance at
$V_{ds}=0$. Here, $\alpha = C_g/C$ is the ratio of gate
capacitance $C_g$ to the total capacitance $C$ and is deduced from the
slopes of the resonance lines in the $V_{ds}-V_g$-plane. The
transformation to $\Delta E$ allows for a direct extraction of
excitation energies from the conductance plot.

\begin{figure}[h]
\begin{center}
\epsfig{file=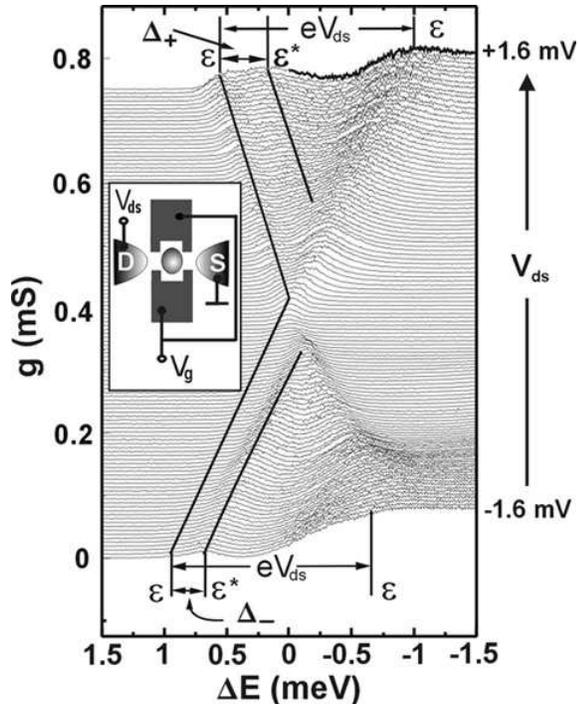,width=7.5cm,keepaspectratio}
\caption{
Bias dependence of the quantum dot conductance in
the vicinity of a single resonance: The drain-source bias is
varied from $V_{ds}=-1.6$~mV (bottom) to $V_{ds}=+1.6$~mV (top) in
120 steps. The horizontal axis is the distance $\Delta E=-e\alpha
\Delta V_g$ from the zero-bias ground state resonance (see text).
The single resonance at $V_{ds}=0$ is split into two resonances
separated by $e V_{ds}$. Additional resonances stem from excited
states of the quantum dot $\Delta_+$ and $\Delta_-$ above the
ground state, for positive and negative bias, respectively. The 
inset shows the gate structure which is used to define the
quantum dot.
}
\end{center}
\end{figure}

From the zero-bias distance between adjacent conductance peaks the
total capacitance of the quantum dot is determined to be $C =
85~$aF. The quantum dot radius is thus estimated to be $r=70~$nm,
i.e.~the quantum dot contains only about 20 electrons. As
expected, for non-zero bias the ground state resonance (marked 
$\epsilon$ in Fig.~1 for comparison with the excited state resonance $\epsilon^*$) 
splits by $eV_{ds}$. For $V_{ds}>0$ an additional conductance resonance due
to an excited state at $\epsilon^*$ develops, which is
$\Delta_+=(\epsilon^*-\epsilon)=390~\mu$eV above the ground state.
Correspondingly, for $V_{ds}<0$ a resonance is detected at a
distance $\Delta_-=280~\mu$eV from the ground state. These
excitation energies are in good agreement with the mean level
spacing $\bar{\Delta} \approx 465~\mu$eV estimated from the dot radius. 
Hence, two different excited states take part in transport 
for $V_{ds}<0$ and $V_{ds}>0$.
Furthermore, as we can see from Fig.~1, the ground state resonance for 
$\Delta E > 0$ is almost suppressed for $V_{ds}<0$, whereas the 
excited state resonance for $\Delta E > 0$ is much stronger.
The origin of these `$\Delta E> 0$'-resonances are the alignment of 
the dot's ground state or the excited state with the chemical  
potential of the drain reservoir. 
The strength of these resonances are related to the overlap 
between the wavefunction of the corresponding quantum dot state 
and the wavefunctions in the reservoirs. 
Hence, the variation in conductance indicates that the coupling of the ground state 
to the drain reservoir is much smaller than that of the excited state. 
This phenomenon was also observed in Ref.~\cite{weis93}.
In our case, we find that the coupling of the excited state to the reservoirs is 
about $5.3$ times the coupling of the ground state.

Two different techniques are applied to study the transport
properties under microwave irradiation. For low forward bias
$V_{ds} \approx 0$, the direct current through the dot is measured
using a single microwave source. Alternatively, we employ two
phase-locked microwave sources which are slightly offset in
frequency. This second technique allows to detect photon-induced
transport also in the nonlinear regime $|V_{ds}|>0$. Furthermore,
the relative phase of the photon-induced current with respect to
the incoming microwave beat can be determined.

\begin{figure}[h]
\begin{center}
\epsfig{file=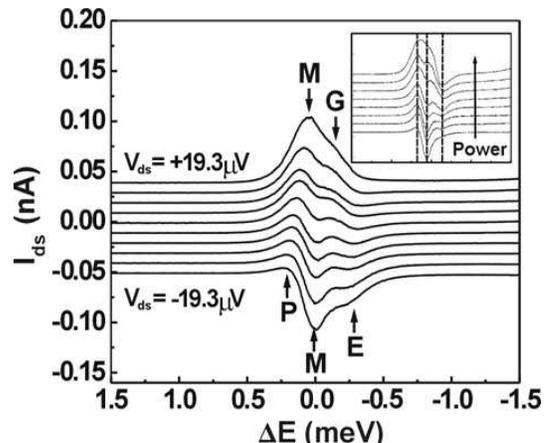,width=7cm,keepaspectratio}
\caption{
Direct current through the quantum dot for small bias
values under irradiation with microwaves of frequency
$f=36.16$~GHz. Next to the main ground state resonance (M) additional
features (G,~P,~E) appear which can be ascribed to photon-assisted tunneling
through the ground state, photon-induced pumping and tunneling 
through an excited state (see text). In the
inset the power dependence of these features is shown for
$V_{ds}=-5.1~\mu$V. The output power of the microwave source is
increased from bottom to top in steps of $0.5$~dBm.
}
\end{center}
\end{figure}

Results obtained with the first technique are shown in Fig.~2, 
where the current through the quantum dot for small bias ranging
from $-19.3~\mu$V to $+19.3~\mu$V, is displayed under microwave
irradiation at frequency $f=36.16~$GHz. To this end, millimeter
waves with frequency $18.08~$GHz are generated by a microwave
synthesizer (HP 87311A), then frequency-doubled (MITEQ MX 2V260400) 
and filtered using a band pass filter (QUINSTAR QFA-3715-BA) 
with center frequency at $32~$GHz. The microwave signal is coupled 
into the cryostat using coaxial lines
and irradiated onto the sample using an antenna formed out of a
conducting loop. The coupling proved to be best at the chosen
frequency $36.16$~GHz. For small positive bias the original main 
peak from the ground state resonance (M) as well as a sideband (G) 
in a distance $h f \approx 0.15~$meV are detected. 
This sideband in the current signal is due
to PAT through the ground state. Quite differently, 
for negative bias additional features in the current
signal are induced by the microwaves. These features can be
attributed to photon-induced pumping (P) and resonant tunneling 
through an excited quantum dot state (E). 
The processes involved are schematically depicted in Fig.~3: 
At low positive bias only the ground state 
transition Fig.~3(a) occurs. As found in the preceding paragraphs, 
the first excited state for this bias direction is too far above the 
ground state to be accessible by a one- or two-photon process. 
The other possible photon-induced ground state transition ($\Delta E > 0$) 
shown in Fig.~3(b) is not detected in the low-bias current signal. 
However, it is resolved for larger bias values applying the two-source detection scheme
(see below). For negative bias $V_{ds}<0$ the excited state at
$\epsilon^*=\epsilon+\Delta_-$ can participate in transport when
the ground state is depopulated by a two-photon absorption process
(Fig.~3(c)) ($2 hf \geq \Delta_-$), a process analogous to
photo-ionization~\cite{oosterkamp97}. 
Normally, this process has a much smaller probability than the one- and two-photon PAT processes. 
In our case, however, since the coupling of the excited state to the reservoirs 
is more than four times stronger than the coupling of the ground state, 
this process might turn out to be comparable to the pure two-photon PAT in amplitude. 
This will be discussed in further detail below.
Furthermore, a pumping current flows opposite to the bias direction for $\Delta E >0$, 
where the ground state is $hf$ above the chemical potentials 
in the source reservoir~(Fig.~3 (d)). 
This only happens when the microwave absorption across the right tunnel 
barrier is larger than that of the left tunneling barrier. 
In this case, the ground state $\epsilon$ is permanently populated with 
electrons from the source contact which then partly decays into the drain region.
From the power dependence (see below), 
we confirm that this pumping current results from PAT. 

Photon-induced features similar to our results have been reported
before~\cite{oosterkamp97} and explained theoretically using, e.g.~nonequilibrium 
Green-function techniques~\cite{stafford96,sun98}. However, to
ensure that the observed features are not adiabatic effects of the
microwave irradiation~(e.g.~rectification effects)~\cite{weis95} 
commonly 
both their frequency and power dependence are determined. 
The inset of Fig.~2 shows the power dependence of the
photon-induced features for $V_{ds}=-5.1~\mu$V. The output power
of the microwave synthesizers is changed in steps of $0.5$~dBm
from trace to trace. Over this wide power range the
microwave-induced features do not change in position showing that
they are indeed induced by single photons. 
We find that the observed dependence of peak heights on microwave power roughly 
agrees with the Bessel function behavior: 
The tunneling current induced by absorbing/emitting $n$ photons is proportional 
to $J^2_n(x)$, where $x=eV_{ac}/hf$ and $V_{ac}$ is the microwave amplitude 
across the tunnel barriers. This behavior was theoretically derived in 
Ref.~\cite{tien63} and was experimentally observed in Ref.~\cite{oosterkamp97,oosterkamp98}. 
For even higher microwave powers the PAT-like features 
considerably broaden due to heating effects until they are finally 
completely washed out. 
Due to the limited bandwidth of our high frequency setup, we are not able to study the freuency 
dependence to identify the photon-induced peaks. Studying the power dependence only is not 
sufficient to reveal the origin of peak~(E). By determining the complex photoconductance, however, 
we will show that the out-of-phase component indicates some aspects of the origin~(see below).

\begin{figure}[h]
\begin{center}
\epsfig{file=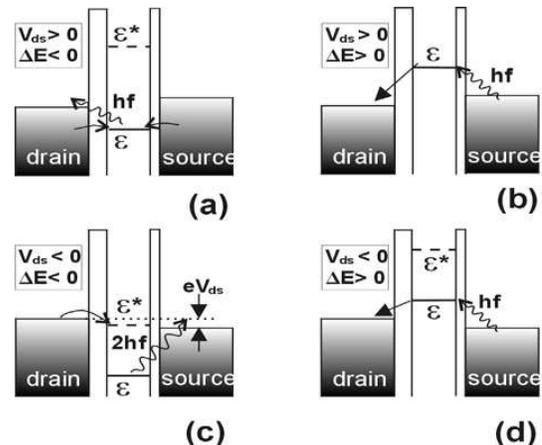,width=7cm,keepaspectratio}
\caption{
Schematic representation of the photon-induced
processes involved in Fig.~2,5,6: In the case of positive bias
(a,~b), only photon-assisted tunneling through the ground state is
possible. For $V_{ds}<0$, the excited state $\epsilon^*$ is
approximately two photon energies above the ground state and can
be accessed by absorption of two photons (c). Furthermore, due to
the asymmetry in microwave absorption across the tunneling barriers, 
a pumping current can occur~(d) for $V_{ds}<0$ (see text).
}
\end{center}
\end{figure}

A more subtle spectroscopic tool applied in this work is the
two-source setup displayed in Fig.~4: Two microwave synthesizers
are phase-locked and tuned to slightly different frequencies
$f_1=18.08~$GHz and $f_2=18.08~$GHz$+\delta f$ with $\delta
f=2.1~$kHz. The two signals are added, frequency-doubled and
filtered with a band pass as described before. Due to the band
pass only microwaves with frequencies $2 f_1$, $f_1+f_2$ and $2
f_2$ are irradiated upon the quantum dot. As these frequency
components have a rigid phase relation, their superposition leads
to a modulated microwave signal with modulation frequency $\delta
f$ (see upper inset of Fig.~4). We have thus produced a flux of
photons with energy $\approx 2 h f_1=0.15$~meV whose intensity
varies periodically in time with frequency $2.1$~kHz. Electronic
transport induced by these photons can be detected with a lock-in
amplifier at the frequency of the microwave beat. Thus, the
detected signal is solely due to the irradiation and contains no
dc contribution. It is therefore possible to observe PAT even in
the non-linear regime, where the broadening of the ordinary
conductance resonances normally masks the photon-induced features.
Another advantage of this technique is the possibility of
heterodyne detection which allows for determination of both
amplitude and relative phase of the signal~\cite{blick98}. This is
not possible using a single microwave source and a simple
modulation technique with a PIN-diode.

\begin{figure}[h]
\begin{center}
\epsfig{file=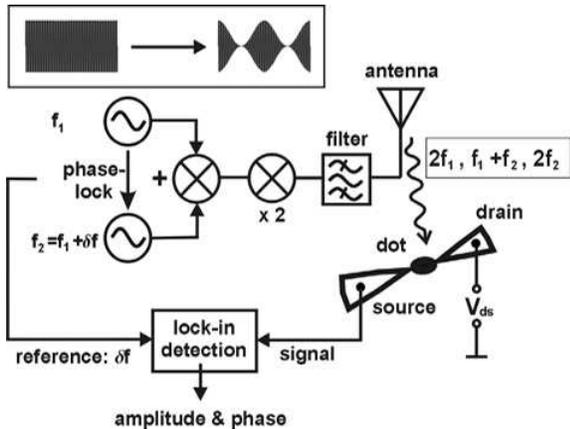,width=7.5cm,keepaspectratio}
\caption{
Experimental setup for the two-source measurement:
Two millimeter waves with a slight frequency-offset generated by
two phase-locked microwave synthesizers ($f1=18.08$~GHz and
$\delta f=2.1$~kHz) are added, doubled and filtered. The resulting
modulated signal is irradiated on the quantum dot by means of an
antenna. With a lock-in amplifier both the amplitude and the phase of
the photoconductance are detected at the modulation frequency
$\delta f$. In the inset the microwave signal before and after
modulation is schematically depicted.
}
\end{center}
\end{figure}

With the lock-in amplifier the in-phase and out-of-phase
photoconductance signals $\gamma_0,\gamma_{\pi/2}$ with respect to
the reference are measured. From these we obtain the total
photoconductance amplitude
$|A|=\sqrt{\gamma_0^2+\gamma_{\pi/2}^2}$ and the relative phase
$\Phi$ which equals $\arctan{(\gamma_{\pi/2}/\gamma_0)}$ for
$\gamma_0 \geq 0$ and $\pi+\arctan{(\gamma_{\pi/2}/\gamma_0)}$ for
$\gamma_0 < 0$, respectively. In Fig.~5 the photoconductance
amplitude at $f=2 f_1=36.16$~GHz and $\delta f=2.1~$kHz is
displayed for the same parameter region as the dc measurement
shown in Fig.~1. With respect to Fig.~1, for $V_{ds}>0$ the
conductance window is enlarged by $2 hf$. The resonances are each
shifted by the photon energy $hf$ which can readily be explained
by photon-assisted tunneling processes as in Fig.~3(a) and (b).
This is also the case for the `$\Delta E > 0$'-conductance
resonances for negative bias. 
However, the resonance for $\Delta E < 0$ and small negative 
bias is clearly shifted by $\Delta_-$, thus enlarging the conductance 
window to $eV_{ds}+hf+\Delta_-$.
The process involved is took as the finite bias version of the transition
depicted in Fig.~3(c): An electron leaves the quantum dot's ground
state for the source reservoir via absorption of two photons. 
Now, electrons can either refill the ground state or tunnel through 
the excited state as long as the ground state is depopulated. 
Transport through the excited state stops when an electron decays to 
the ground state, or an electron enters the quantum dot's ground state from the leads.
With $\Delta E < 0$ and larger negative bias, the photoconductance peak is apparently broadened. 
The broadening is partly due to other tunneling processes possible at large bias, 
e.g.~one-photon PAT through the ground state. In fact, even at small negative bias 
there is small tunneling current in between the peak~(M) and (E), 
which is most probably the one-photon PAT. In our case, the tunneling 
process for $\Delta E < 0$ and negative bias is more intricate than the ideal PAT.

\begin{figure}[h]
\begin{center}
\epsfig{file=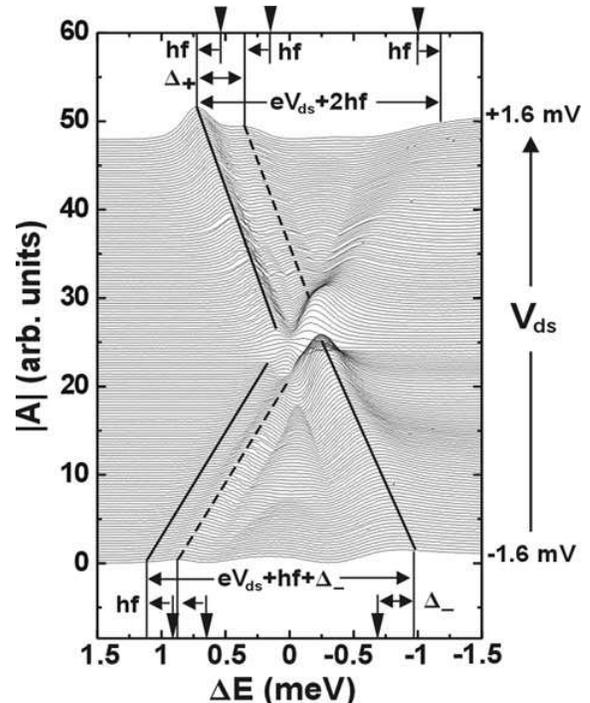,width=7.5cm,keepaspectratio}
\caption{
Amplitude $|A|$ of the photoconductance
measurement obtained with the two-source setup of Fig.~4. The
drain-source bias is varied as in Fig.~1. The position of the
resonances found in the dc measurement of Fig.~1 are indicated
with triangles. With respect to the dc resonances, most
photoconductance resonances are shifted by the photon energy $h
f$, as can be expected for photon-assisted processes. The
resonance for $V_{ds}<0$ and $\Delta E<0$, however, is shifted by
$\Delta_-$, indicating two-photon PAT through a ground state and 
the resulting tunneling through an excited state as sketched in Fig.~3~(c).
}
\end{center}
\end{figure}

\begin{figure}[h]
\begin{center}
\epsfig{file=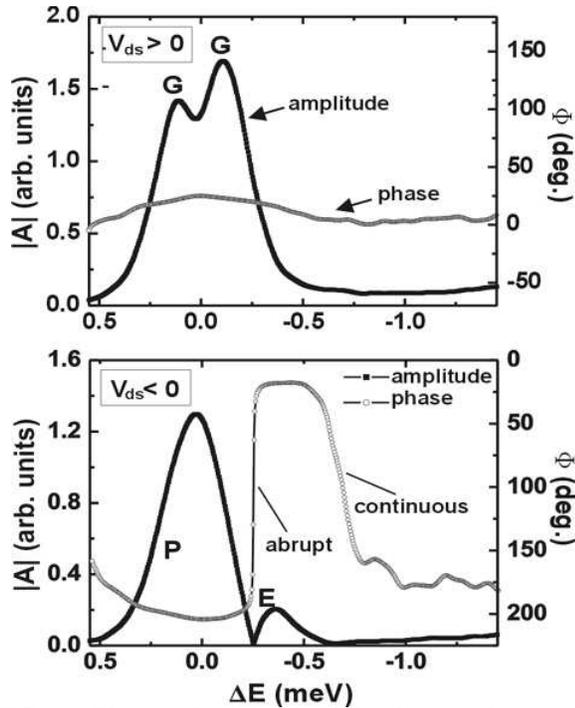,width=7.5cm,keepaspectratio}
\caption{
Phase and amplitude of the photoconductance signal
for $V_{ds}=+10~\mu$V (top) and $V_{ds}=-30~\mu$V (bottom). For
positive bias, photon-assisted tunneling through the ground state
(G) is observed, as schematically depicted in Fig.~3~(a) and (b).
The phase signal remains constant on either side of the
resonances. For negative bias, pumping (P) and tunneling through
an excited state (E) are induced (cf. Fig.3~(d) and (c)). Vanishing
of the amplitude signal between (P) and (E) is accompanied by a
trivial jump of $\pi$ in the phase signal. Finally, for more
negative values of $\Delta E$  the phase signal continuously falls
back to its original value.
}
\end{center}
\end{figure}

In Fig.~6, phase traces as well as their respective amplitude
signals are displayed for small positive and negative bias
($V_{ds}=+10~\mu$V and $V_{ds}=-30~\mu$V, respectively),
corresponding to the central region of Fig.~5. For $V_{ds}>0$ the
phase signal remains approximately constant at $\Phi=0$ which
means that the out-of-phase photoconductance $\gamma_{\pi/2}$ is
equal to zero. The response of the quantum dot to the microwaves
is similar for both of the tunneling processes (G). In fact, the
two peaks (G) in the amplitude signal stem from ground state
resonances as depicted in Fig.~3(a) and (b). The situation is
considerably different for $V_{ds} < 0$ where a strong pumping
signal (P) is observed which is caused by a process as in
Fig.~3(d). At the position where the photocurrent changes its
direction, the amplitude drops to zero and the phase changes
trivially by $\pi$ (this corresponds to crossing zero in the
$\gamma_0-\gamma_{\pi/2}$-plane). The second peak~(E) stems from 
the photon-induced tunneling through the excited state as in Fig.~3(c). 
Moving away from this second resonance to more negative $\Delta E$, 
the phase continuously returns to its original value.

This continuous phase change shows that this transport process
results in a finite out-of-phase signal $\gamma_{\pi/2}$. In
contrast to the other transport scenarios described above~(only the ground state is involved),
photon-induced tunneling through the excited state is not a purely
conductive transport process but also has capacitive and inductive
contributions. This behavior is due to the complicated 
charging dynamics of the quantum dot for this particular process. 
The processes involved are PAT from the ground state to the source 
reservoir, resonant tunneling through the excited state, 
recharging of the ground state by the drain reservoir and 
relaxation from the excited state to the ground state. 
All these processes have different time constants which additionally 
depend on the gate voltage (i.e.~$\Delta E$). 
The interplay of these processes results in the observed phase lag. 
Thus one has a method at hand to determine the admittance of 
a mesoscopic system~\cite{fu93,buttiker93a,buttiker93b} in the PAT 
regime which is related to the average relaxation time of the system. 
In the current setup, for $V_{ds}<0$ the ground state broadening, 
due to the coupling to the drain reservoir, is about $400$~MHz, 
while the level broadening from the coupling to the source reservoir is around $2$~GHz. 
The broadening of the excited state coupling to the reservoirs is 
found to be of the same width of $2$~GHz. 
Hence, the bare tunneling time through the ground state, 
excluding other time constants, would be less than $2.5$~ns. 
However, the inverse modulation frequency $1/\delta f \approx 500~\mu s$, 
which is the time separation between two microwave beat minima, 
is much larger than the tunneling time. 
In the few-electron limit, this indicates that it takes the electron 
a much longer time to relax within the dot than to tunnel through the barriers. 
An extension of the measurements to modulation frequencies
on the order of $10-100~$MHz corresponding to a time scale of
$10-100~$ns would therefore be desirable. 
With a shorter microwave beat period we will be able to probe both the fast tunneling event 
and the slow relaxation process. 
We conclude that with the frequency $f$ the photon energy $hf$ for 
the photon-induced process can be adjusted, whereas the modulation 
frequency $\delta f$ determines the time scale on which the electronic 
dynamics of the quantum dot is probed.

In summary, 
we have presented complex photoconductance measurements in the non-linear
transport regime of a few-electron quantum dot using phase-locked
microwave sources. The electronic structure of the dot is first
characterized by conventional conductance measurements without microwave
radiation. Photon-assisted tunneling through the ground state as
well as through excited states of the system is observed. 
The two-source method allows to perform PAT measurements even in the
non-linear transport regime. 
Most importantly, the relative phase of the photocurrent with respect to 
the incoming microwave beat signal can be obtained from the two-source measurement. 
This phase is related to the susceptance of the quantum dot at very high frequencies.
Non-trivial values for this quantity can be attributed to the long charge relaxation
times in the quantum dot. In future work this can be exploited for an accurate
determination of the relaxation times of excited quantum dot states.
\\

We like to thank Q.~F.~Sun, A.~W.~Holleitner
and S.~Manus for helpful discussions. This work was funded in part
by the Deutsche Forschungsgemeinschaft within project SFB~348  
and the Defense Advanced Research Projects Agency (DARPA) 
Ultrafast Electronics Program. H.~Q. gratefully acknowledges 
support by the Volkswagen Stiftung.\\

$\dag$: present address: Bell Laboratories, Lucent Technologies,
600 Mountain Ave, Murray Hill, NJ 07974, USA

$\ddag$ new address: Universit\"at Regensburg, Universit\"atsstr.~31,
D-93040 Regensburg, Germany.



\begin{thebibliography}{10}
\bibitem{sloppy}
For simplicity, we adopt a rather sloppy formulation here. In
fact, it is the discrete chemical potential $E(N+1)-E(N)$ of the
quantum dot associated with the ground state transition $N+1
\leftrightarrow N$ which is aligned with the chemical potential of
the leads.
\end{thebibliography}
\end{document}